# HYBRIDS : DYNAMICS AND DISGUISES

P.R. PAGE

*Theoretical Physics, University of Oxford,*
*1 Keble Road, Oxford OX1 3NP, UK*

We introduce selection rules arizing from flux–tube dynamics with non–relativistic and adiabatic quark motion. Specifically, we indicate how states can disguise themselves by not decaying to S–wave states.

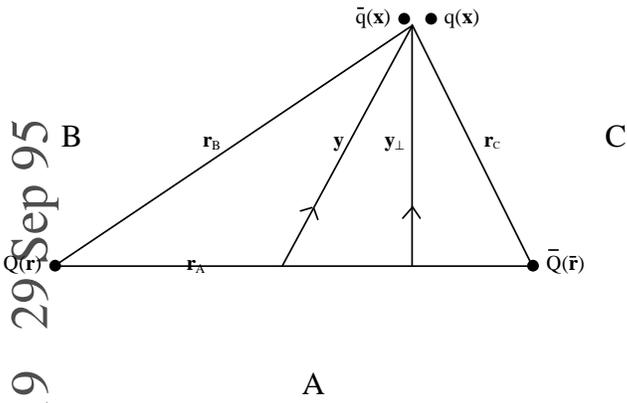

Mesons can be regarded as $Q\bar{Q}$ systems connected by a flux–tube. As a first orientation we conceptually seperate the dynamics of the "quark" and "flux" components of the system. This is called the *adiabaticity* assumption, and is valid for large quark masses. As the quarks move, the flux–tube is considered to re-assemble itself spontaneously. Hybrids are $Q\bar{Q}$ systems with an excited flux–tube. The flux–tube can possess angular momentum $\Lambda_H$ around the $Q\bar{Q}$–axis $\mathbf{r}_H$, which is conserved as the flux–tube $\mid \mathbf{r}_H H \rangle$ must *a priori* be invariant under rotations around the $Q\bar{Q}$–axis. The quarks move adiabatically in an effective potential generated by the flux–tube dynamics, and hence obey a Schrödinger equation which determines the dependence on $Q\bar{Q}$ seperation of the wave function $\psi_H(\mathbf{r}_H)$ of the system. The above ideas for mesons and hybrids have been implemented in lattice gauge theory[1] and in simulations thereof[2].

We proceed to discuss the decay dynamics of states with adiabatically and non–relativistically[1] moving quarks. The term "states" refers to *both* mesons and hybrids. We assume that a $q\bar{q}$ pair is created with quark mass $m$ at position $\mathbf{y}$ with spin $S_{q\bar{q}} = 1$, orbital angular momentum $L_{q\bar{q}} = 1$ and total angular momentum $J_{q\bar{q}} = 0$ (see Figure for the decay topology). The decay process is called "$^3P_0$ pair creation". Its pre–eminence as a decay model for mesons is based on its surprising phenomenological success[3] especially for light mesons[2], in contrast to $^3S_1$ pair creation for example. We adopt it.

Given the success of $^3P_0$ pair creation for meson decays to mesons, we extrapolate this mechanism to all states. We shall develop selection rules valid for arbitrary wave functions. By the conservation of spin (since the decay operator creates a $S_{q\bar{q}} = 1$ pair) we firstly obtain the following selection rule :

*Decays of net spin $S = 0$ states to two $S = 0$ states are forbidden.*

The $^3P_0$ pair creation amplitude can be shown[2,4] to be proportional to

$$\int d^3\mathbf{r}_A \, d^3\mathbf{y} \, \psi_A(\mathbf{r}_A) \, \exp(i\frac{M}{m+M}\mathbf{p}_B \cdot \mathbf{r}_A) \, \gamma(\mathbf{r}_A, \mathbf{y})$$
$$\times (i\boldsymbol{\nabla}_{\mathbf{r}_B} + i\boldsymbol{\nabla}_{\mathbf{r}_C} + \frac{2m}{m+M}\mathbf{p}_B) \, \psi_B{}^*(\mathbf{r}_B) \, \psi_C{}^*(\mathbf{r}_C) \quad (1)$$

for a stationary state A with quarks of mass $M$ decaying to the outgoing states B and C. The quark and flux degrees of freedom are seperated adiabatically. The initial flux–tube would have to re–assemble into the two flux–tubes of the final states. This has a certain re–arrangement amplitude, which we call the *flux–tube overlap* $\gamma(\mathbf{r}_A, \mathbf{y})$.

The pair creation position $\mathbf{y}$ can be decomposed into the transverse "component" $\mathbf{y}_\perp \equiv -(\mathbf{y} \times \hat{\mathbf{r}}_A) \times \hat{\mathbf{r}}_A$ and the parallel "component" $\mathbf{y}_\parallel \equiv (\mathbf{y} \cdot \hat{\mathbf{r}}_A)\hat{\mathbf{r}}_A$ (with magnitude $y_\parallel \equiv \mathbf{y} \cdot \hat{\mathbf{r}}_A$). Defining $\phi$ as the angle around the $Q\bar{Q}$–axis, we introduce a result related to the conservation of angular momentum around the $Q\bar{Q}$–axis.

**Theorem 1** *The most general form of the flux–tube overlap in the limit where the pair creation is near to the initial $Q\bar{Q}$-axis is*

$$\gamma(\mathbf{r}_A, \mathbf{y}) = e^{i\Lambda\phi} \, f(r_A, \mathbf{y}_\perp^2, y_\parallel) \qquad where \quad \Lambda \equiv \Lambda_A - \Lambda_B - \Lambda_C \quad (2)$$

**Proof** The full decay configuration can be described[2] by the six variables $\mathbf{r}_A, \mathbf{y}$. By rotational invariance the overlap $\gamma(\mathbf{r}_A, \mathbf{y})$ cannot depend on the direction $\hat{\mathbf{r}}_A$ of the $Q\bar{Q}$–axis. This leaves the dependence to be on the four variables $r_A, \mathbf{y}_\perp^2, \phi$ and $y_\parallel$. We can reveal the $\phi$–dependence by considering a rotation $\mathcal{R}$ of an initial pair creation position $\mathbf{y}$ corresponding to $\phi = 0$ by an angle $\phi$ around the $Q\bar{Q}$–axis. Denoting the effect of pair creation by $\hat{O}$

$$\gamma(\mathbf{r}_A, \mathcal{R}\mathbf{y}) \equiv \langle \mathbf{r}_B B \, \mathbf{r}_C C | \, \hat{O}(\mathcal{R}\mathbf{y}) \, | \mathbf{r}_A A \rangle$$
$$= \langle \mathbf{r}_B B \, \mathbf{r}_C C | \, \mathcal{R}^+\hat{O}(\mathbf{y})\mathcal{R} \, | \mathbf{r}_A A \rangle = \langle \mathcal{R}\mathbf{r}_B B \, \mathcal{R}\mathbf{r}_C C | \, \hat{O}(\mathbf{y}) \, | \mathcal{R}\mathbf{r}_A A \rangle \quad (3)$$

giving the desired result since $\mathcal{R} \, |\mathbf{r}_H H\rangle = \exp(i\phi\Lambda_H) \, |\mathbf{r}_H H\rangle$, where $H \in \{A, B, C\}$. □

We note that when the y–integral in Eq. 1 is performed the $\phi$–dependence $\exp i\Lambda\phi$ in the flux–tube overlap must be matched by a factor $\exp -i\lambda\phi$ arising from the y–dependent part of the decay amplitude which only contributes when $\lambda = \Lambda$ using $\int_0^{2\pi} d\phi \, e^{-i\lambda\phi}e^{i\Lambda\phi} = 2\pi\delta_{\lambda\Lambda}$

**Theorem 2** *For pair creation near the initial $Q\bar{Q}$-axis, decay is forbidden for (1) $|\Lambda_A| \geq 2$ state $\rightarrow$ two S–wave states; (2) $|\Lambda_A| = 1$ state $\rightarrow$ two identical S–wave states, if $\gamma(\mathbf{r}_A, \mathbf{y})$ is even under $y_\parallel \rightarrow -y_\parallel$; (3) $|\Lambda_A| = 0$ state $\rightarrow$ two identical S–wave states, if $\gamma(\mathbf{r}_A, \mathbf{y})$ is odd under $y_\parallel \rightarrow -y_\parallel$.*

**Proof** Because $\psi_H(\mathbf{r}_H)$, $H \in \{B, C\}$, is an S–wave wave function it only depends on $\mathbf{r}_H{}^2$, i.e. $\psi_H(\mathbf{r}_H) \equiv \tilde{\psi}_H(\mathbf{r}_H^2)$, and hence in the last line in Eq. 1 on the $\phi$–independent variable $\mathbf{r}_H{}^2 = \mathbf{r}_A^2/4 \pm \mathbf{r}_A.\mathbf{y} + \mathbf{y}^2$ when we substitute $\mathbf{r}_B = \mathbf{r}_A/2 + \mathbf{y}$, $\mathbf{r}_C = \mathbf{r}_A/2 - \mathbf{y}$. The non–derivative term in Eq. 1 has $\lambda = 0$ due to its $\phi$–independence. Moreover, the derivative terms in the last line of Eq. 1 equals

$$2i \left\{ (\frac{\mathbf{r}_A}{2} + \mathbf{y}) \nabla_{\mathbf{r}_B{}^2} \tilde{\psi}_B^*(\mathbf{r}_B^2) \, \tilde{\psi}_C^*(\mathbf{r}_C^2) + (\frac{\mathbf{r}_A}{2} - \mathbf{y}) \, \tilde{\psi}_B^*(\mathbf{r}_B^2) \nabla_{\mathbf{r}_C{}^2} \tilde{\psi}_C^*(\mathbf{r}_C^2) \right\} \quad (4)$$

by the chain rule. In the above $\lambda = -1, 0, 1$ because $\nabla_{\mathbf{r}_B{}^2} \tilde{\psi}^*(\mathbf{r}_B^2) \, \tilde{\psi}_C^*(\mathbf{r}_C^2)$ and $\tilde{\psi}_B^*(\mathbf{r}_B^2) \nabla_{\mathbf{r}_C{}^2} \tilde{\psi}^*(\mathbf{r}_C^2)$ are $\phi$–independent. So clearly if $|\Lambda| \geq 2$ there is no contribution, establishing the first result. For the remaining results define the common wave function $\psi \equiv \psi_B = \psi_C$. The last line of Eq. 1 equals

$$(2i\nabla_{\mathbf{r}_A} + \frac{2m}{m+M}\mathbf{p}_B) \, \psi^*(\frac{\mathbf{r}_A}{2} + \mathbf{y})\psi^*(\frac{\mathbf{r}_A}{2} - \mathbf{y}) \quad (5)$$

If $|\Lambda| = 1$ Eq. 2 implies that $\gamma(\mathbf{r}_A, \mathbf{y})$ is odd under $\mathbf{y}_\perp \rightarrow -\mathbf{y}_\perp$, and by assumption it is even under $y_\parallel \rightarrow -y_\parallel$, so that it is odd under exchange of $\mathbf{y} \rightarrow -\mathbf{y}$. Analogously, if $|\Lambda| = 0$, $\gamma(\mathbf{r}_A, \mathbf{y})$ is even under $\mathbf{y}_\perp \rightarrow -\mathbf{y}_\perp$, and by assumption odd under $y_\parallel \rightarrow -y_\parallel$, so that it is odd under $\mathbf{y} \rightarrow -\mathbf{y}$. It is sufficient to show that Eq. 5 is even under $\mathbf{y} \rightarrow -\mathbf{y}$. But this is manifest since the derivative term in Eq. 5 is independent of $\mathbf{y}$ and $\psi^*(\frac{\mathbf{r}_A}{2} + \mathbf{y}) \, \psi^*(\frac{\mathbf{r}_A}{2} - \mathbf{y})$ is symmetric under $\mathbf{y} \rightarrow -\mathbf{y}$. $\square$

We expect (partial) breaking of the above rule when the outgoing states do not have identical wave functions, making the decay amplitude proportional to the "difference of the final state wave functions". Theorem 2 also has the consequence that modes not suppressed by this rule might in the absence of arguments to the contrary be considered potentially significant.